# Anomalous narrow-band correlation in a natural superconducting heterostructure

Xiupeng Sun[1#], Zhiyuan Wei[1#], Min Shan[1], Shuting Peng[1], Yang Luo[1], Jianchang Shen[1], Linwei Huai[1], Yu Miao[1], Zhipeng Ou[1], Mehmet Onbasli[2], Zhenyu Wang[1], Tao Wu[1], Junfeng He[1]*, Xianhui Chen[1]*

[1]*Department of Physics and CAS Key Laboratory of Strongly-coupled Quantum Matter Physics, University of Science and Technology of China, Hefei, Anhui, 230026, China*
[2]*Department of Electrical and Electronics Engineering and Department of Physics, Koç University, Istanbul, 34450, Türkiye*

[#]These authors contributed equally to this work
*To whom correspondence should be addressed: jfhe@ustc.edu.cn, chenxh@ustc.edu.cn

**A new frontier in condensed matter physics is to stack atomically thin layered-materials with different properties and create intriguing phenomena which do not exist in any of the constituent layers. Transition metal dichalcogenide 4Hb-TaS$_2$, with an alternating stacking of a spin liquid candidate 1T-TaS$_2$ and a superconductor 1H-TaS$_2$, is a natural heterostructure for such a purpose. Recently, rare phenomena are indeed observed, including chiral superconductivity, two-component nematic superconductivity, topological surface superconductivity and enigmatic magnetic memory. A widely proposed starting point to understand such a mysterious heterostructure requires strong electronic correlation, presumably provided by 1T-TaS$_2$ layers with a narrow flat band near the Fermi level ($E_F$). Here, by using angle-resolved photoemission spectroscopy, we reveal the theoretically expected flat band near $E_F$ in the energy-momentum space for the first time. However, this flat band only exists on the 1T-TaS$_2$ terminated surface layer with broken translational symmetry, but not on the 1T-TaS$_2$ layers buried in the bulk. These results directly challenge the foundation of the current theoretical paradigm. On the 1T-TaS$_2$ terminated surface layer, we further reveal a pseudogap and an anomalous doping effect. These phenomena and the dichotomy between surface and bulk layers also shed new light on the unusual coexistence of distinct electronic orders in this mysterious heterostructure.**

In flat band material systems, when the Fermi level ($E_F$) is aligned with the flat band, the kinetic energy of electrons is strongly suppressed and a moderate Coulomb interaction between electrons would become significantly greater than the kinetic energy. As such, the system is driven into a correlated electronic state and nontrivial electronic orders may emerge. For instance, the flat band in twisted bilayer graphene is essential for the electronic correlation and superconductivity[1,2]. In transition metal dichalcogenide (TMD) 4Hb-TaS$_2$ heterostructure[3-19], a charge density wave (CDW) with periodic Star-of-David (SD) clusters is formed on the 1T-TaS$_2$ layer at low temperature, giving rise to a flat band and a CDW energy gap[3-6,17,20]. This flat band is theoretically predicted near E$_F$, providing strong electronic correlation to the material sysmtem[12,17,18]. Scanning tunneling spectroscopy measurements reveal a sharp peak in electron density of states at an energy position slightly above E$_F$[5,6,19,21], being consistent with that of a flat band. However, direct experimental evidence of the expected flat band in the energy-momentum space is still lacking. Here, we investigate this issue by taking advantage of an angle-resolved photoemission spectroscopy (ARPES) system with high spatial resolution. As such, the electronic structures on different terminations of the 4Hb-TaS$_2$ heterostructure are examined separately.

As shown in Fig. 1a-d, when the alternately stacked 4Hb-TaS$_2$ single crystal is cleaved, either 1H-TaS$_2$ (Fig. 1b) or 1T-TaS$_2$ (Fig. 1d) terminated layer appears on the top (also see supplementary Fig. S1). Using a small beam spot, termination dependent ARPES measurements are performed at low temperature. As shown in Fig. 1e, Ta 4$f_{7/2}$ core-level spectrum measured on the 1H termination exhibits a dominant peak at around 22.7 eV below E$_F$. The same measurement on the 1T termination displays two dominant peaks at around 23.3 eV and 23.8 eV below E$_F$, respectively (Fig. 1f). These results are consistent with the X-ray photoelectron spectroscopy (XPS) studies on 2H-TaS$_2$[22,23] and 1T-TaS$_2$[24], respectively. Fermi surface measurement on the 1H termination reveals circular hole pockets around Γ and K points, and dog-bone like electron pocket around M point (Fig. 1g), being consistent with the Fermi surface sheets observed on bulk crystals of 2H-TaS$_2$[25,26]. On the other hand, the Fermi surface measurement on the 1T termination shows significant spectral weight near Γ point (Fig. 1h), bearing a resemblance to the Fermi surface mapping measured on the bulk 1T-TaS$_2$ at low temperature[27-29]. Strikingly, features of the 1H termination (circular hole pockets and dog-bone like electron pockets) are also discernible on the 1T termination (Fig. 1h).

There are two possible scenarios for the above paradox. The first one is to consider a trivial contamination from the 1H termination (Fig. 2a). In this situation, the photon beam illuminates two terminations simultaneously and electronic structures from both terminations are probed and simply combined in the experimental data. The second scenario is to consider photoelectrons from both the 1T (1H) surface layer and the 1H (1T) layer beneath it, for the measurements on the 1T (1H) termination (Fig. 2b). In this case, the 1T (1H) surface layer might not be equivalent to that in the bulk due to the broken translational symmetry, and the corresponding electronic structure may also change. To resolve this issue, termination dependent measurements of the band structure are carried out. Two sets of bands are observed on both terminations (Fig. 2d,e), showing the characteristic features of the H-phase[26,30] and T-phase[27,31-34], respectively. First, on the 1H termination, electron energy bands of the 1H layer exhibit dominate intensity (Fig. 2d, marked by black dashed lines) with multiple Fermi crossings at $E_F$. Electron energy bands of the 1T layer are also observed but with weaker intensity (Fig. 2d, marked by red dashed lines). These 1T bands are below $E_F$ without any Fermi crossing. Such results are consistent with the Fermi surface mapping on the 1H termination (Fig. 1g), where the Fermi surface is completely formed by the bands of 1H layer. Second, on the 1T termination, electron energy bands of the 1H layer are still discernible, but with much weaker intensity (Fig. 2e, black dashed lines). Bands of the 1T layer are now probed with significantly enhanced intensity (Fig. 2e, red dashed lines). Strikingly, the energy locations of the 1T bands exhibit a substantial shift towards a deeper binding energy and a flat band is resolved near $E_F$ (Fig. 2e, marked by pink boxes). This flat band is better visualized when the Fermi Dirac function is removed (Fig. 3a), revealing its energy location at ~40 meV above $E_F$. We note that the flat band and the energy gap between the flat band and the valence band top are fully consistent with theoretical predictions of the 1T-CDW state[12,18], and are experimentally revealed in the energy-momentum space for the first time. These results demonstrate that the two sets of bands observed on the 1H termination and 1T termination are nonequivalent, featuring the second scenario as illustrated in Fig. 2b. In this context, it is important to point out that the theoretically expected flat band near $E_F$ only exists on the 1T-TaS$_2$ terminated surface layer with broken translational symmetry (Fig. 2e). On the contrary, the flat band of the buried 1T-TaS$_2$ layer (Fig. 2d) is well above $E_F$, which cannot be probed by photoemission measurements (compare Fig. 2d with Fig. 2e).

Then we examine temperature evolution of the 1T-CDW state and the associated flat band on the 1T-TaS$_2$ terminated surface layer (Fig. 3). The CDW transition temperature of our sample is ~327 K (supplementary Fig. S2), providing an opportunity to reveal the energy bands slightly above $E_F$ by thermal population. In the CDW state (150 K), the flat band near $E_F$ and the energy gap between the flat band and the valence band top (Fig. 3a) can be quantified by raw energy distribution curves (EDCs) in Fig. 3d,e. With increasing temperature, the valence band moves towards $E_F$ and electron spectral weight starts to fill into the energy gap (Fig. 3a-e). Surprisingly, the flat band and CDW energy gap are still discernible at elevated temperatures above the CDW transition temperature (Fig. 3b,c,e,f), featuring the existence of a CDW pseudogap (also see supplementary Fig. S3). In the meantime, the energy location of the flat band shows little change with temperature. These results point to unconventional behaviors of the CDW state.

Next, we explore doping dependence of the 1T-CDW state on the 1T-TaS$_2$ terminated surface layer, by *in-situ* potassium (K) deposition (Fig. 4). Previous studies on 1T-TaS$_2$ single crystals reveal a sudden change to band insulator with K doping[35-37]. In that case, each K atom is adsorbed at the center of a SD cluster[35-37], and the electron donated from K would form a strong local bonding with the single electron of the SD cluster, resulting in a band insulator with a full-filled band well below $E_F$ (Fig. 4a). In the meantime, the electronic structure of other SD clusters without K adsorbate remains unaffected[36,37]. Whether such a two-component local picture can describe our current system is examined by continuous K deposition and ARPES measurements. As shown in Fig. 4c,f, initial K surface doping gives rise to a nondispersive band at ~0.12 eV below $E_F$, whereas the flat band near $E_F$ and the CDW gap remain unchanged. These results seem to be consistent with the two-component local picture reported in 1T-TaS$_2$ [36,37]. Nevertheless, further K doping leads to a substantial change of the electronic structure (Fig. 4d,g), where the nondispersive new band moves towards a deeper binding energy, but the original hole-like band shifts towards $E_F$. The opposite energy shift of different bands indicates a new physical process beyond the two-component local picture.

Finally, we discuss the implications of our experimental observations. First, electronic structures on different terminations of 4Hb-TaS$_2$ are distinguished and the theoretically predicted flat band is indeed observed near $E_F$ in the energy-momentum space for the first time, but only on the 1T surface layer. Therefore, a direct question is to understand the dichotomy between the 1T surface layer and the 1T layer buried in the bulk. The electronic structures of these two types of 1T layers exhibit similar band dispersion but with a significant energy shift, indicating a substantial difference in electron occupation. A plausible explanation is to consider the charge transfer between different layers. It has been reported that electrons would naturally transfer from 1T layers to 1H layers in the 4Hb-TaS$_2$ system[19,38]. When the 1T layer is on the surface, only one adjacent 1H layer is attracting electrons from the 1T layer (Fig. 2c). However, when the 1T layer is in the bulk, two adjacent 1H layers are extracting electrons from the 1T layer (Fig. 2c). Therefore, the 1T surface layer is less hole-doped comparing to that in the bulk, being consistent with the experimental results. In this context, we note that the 1H layer exhibits multiple electron-like bands crossing $E_F$. As such, a moderate energy shift of these bands is induced by the same amount of carrier difference between the surface and bulk. Second, it is widely believed that the 1T layers in 4Hb-TaS$_2$ provide strong electronic correlation to the material system by a narrow flat band near $E_F$. Our results demonstrate that such a mechanism only works for the 1T layer on the surface, but not for the 1T layers in the bulk material. In this context, it would be interesting to explore whether the different electronic correlation between the surface and bulk is relevant to the reported multi-component electronic orders and the dichotomy between surface and bulk properties in this material[8-11,15]. Third, the persistence of the energy gap and flat band at elevated temperatures above the CDW transition temperature indicates the existence of incoherent short-range CDW orders. This is consistent with the broadened spectral peak at high temperature (Fig. 3e). More theoretical efforts are stimulated to investigate the flat band and the associated electronic correlation in short-range CDW state. Fourth, the continuous K surface doping on the 1T surface layer of 4Hb-TaS$_2$ reveals new physics beyond the traditional two-component local picture. The most striking observation is the opposite energy shift between the nondispersive new band and the original hole-like band. A possible scenario is to consider itinerant electrons of K atoms. On one hand, the electron doping pushes the nondispersive new band towards a deeper binding energy. On the other hand, these itinerant electrons enhance the screening effect which reduces the CDW gap. As such, the original hole-like band shifts toward $E_F$. We note that the traditional two-component local picture has been serving as a basis to understand the band insulating phase and Mott insulating phase in 1T-$TaS_2$ [36,37]. The current observations in K surface doped 4Hb-$TaS_2$ reveal the important role of itinerant electrons beyond the local physics.

In conclusion, our termination dependent ARPES measurements reveal the theoretically expected flat band near $E_F$ in the energy-momentum space, but only on the 1T surface layer. Temperature and doping dependent measurements further unveil new phenomena on the 1T surface layer beyond the standard paradigm. These observations directly challenge the current understanding of the 1T-$TaS_2$ building block and provide a new route to examine the mysterious electronic orders in the 4Hb-$TaS_2$ heterostructure.

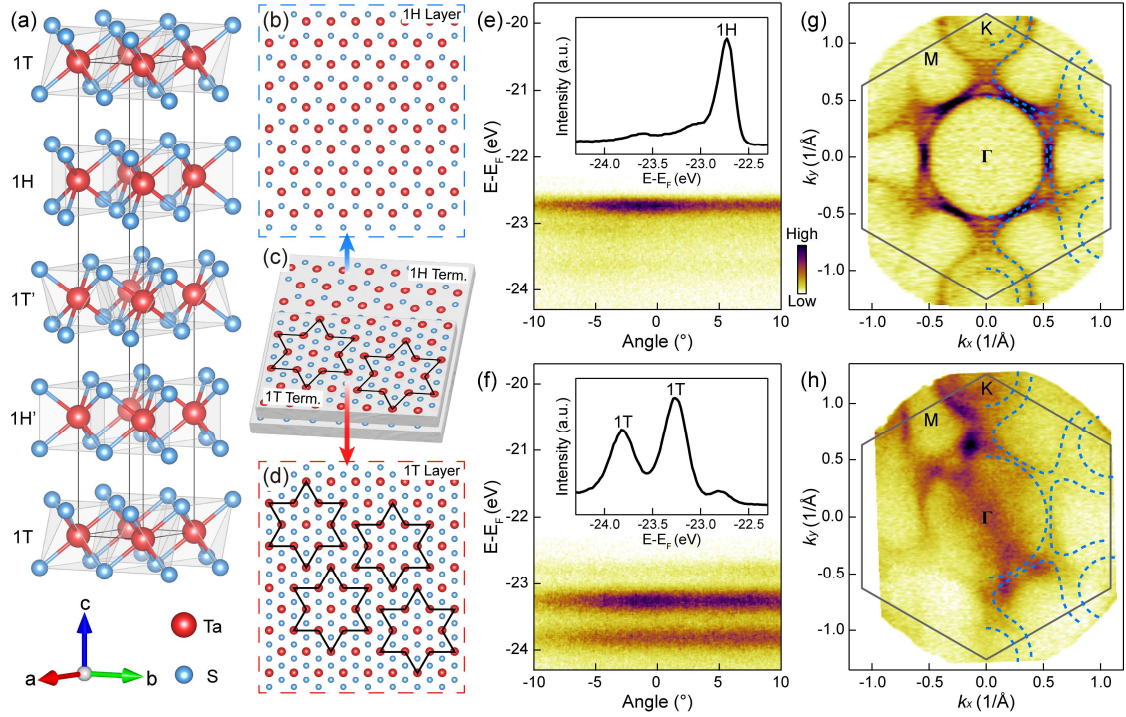

**Fig. 1 Crystal structure and termination dependent measurements of the Fermi surface at 150K. a,** Crystal structure of 4Hb-TaS$_2$. **b,** Top view of the 1H layer. **c,** Schematic of the two terminations of 4Hb-TaS$_2$. **d,** Top view of the 1T layer with periodic SD clusters in the CDW state. **e,f,** Ta $4f_{7/2}$ core-level spectrum measured on the 1H and 1T termination, respectively. The integrated EDC of the core-level spectrum is shown in the inset. **g,h,** Fermi surface measured on the 1H and 1T termination, respectively. Blue dashed lines are a guide to the eye.

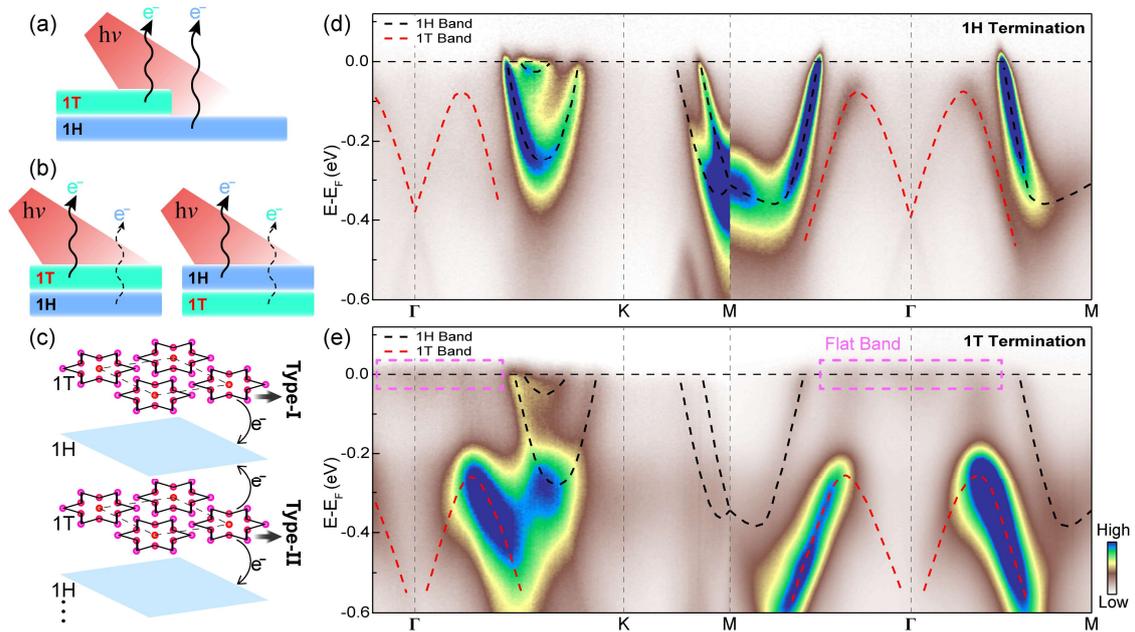

**Fig. 2 Termination dependent measurements of the energy bands at 150K. a,b,** Schematic of possible scenarios for the measured photoelectrons. **c,** Schematic of two types of 1T layers and the different amount of charge transfer to the adjacent 1H layer(s). **d,e,** Photoemission intensity plot along Γ-K-M-Γ-M of the 1H and 1T termination, respectively. The black and red dashed lines mark the 1H-bands and 1T-bands, respectively. The pink dashed boxes in **e** indicate the flat band near $E_F$.

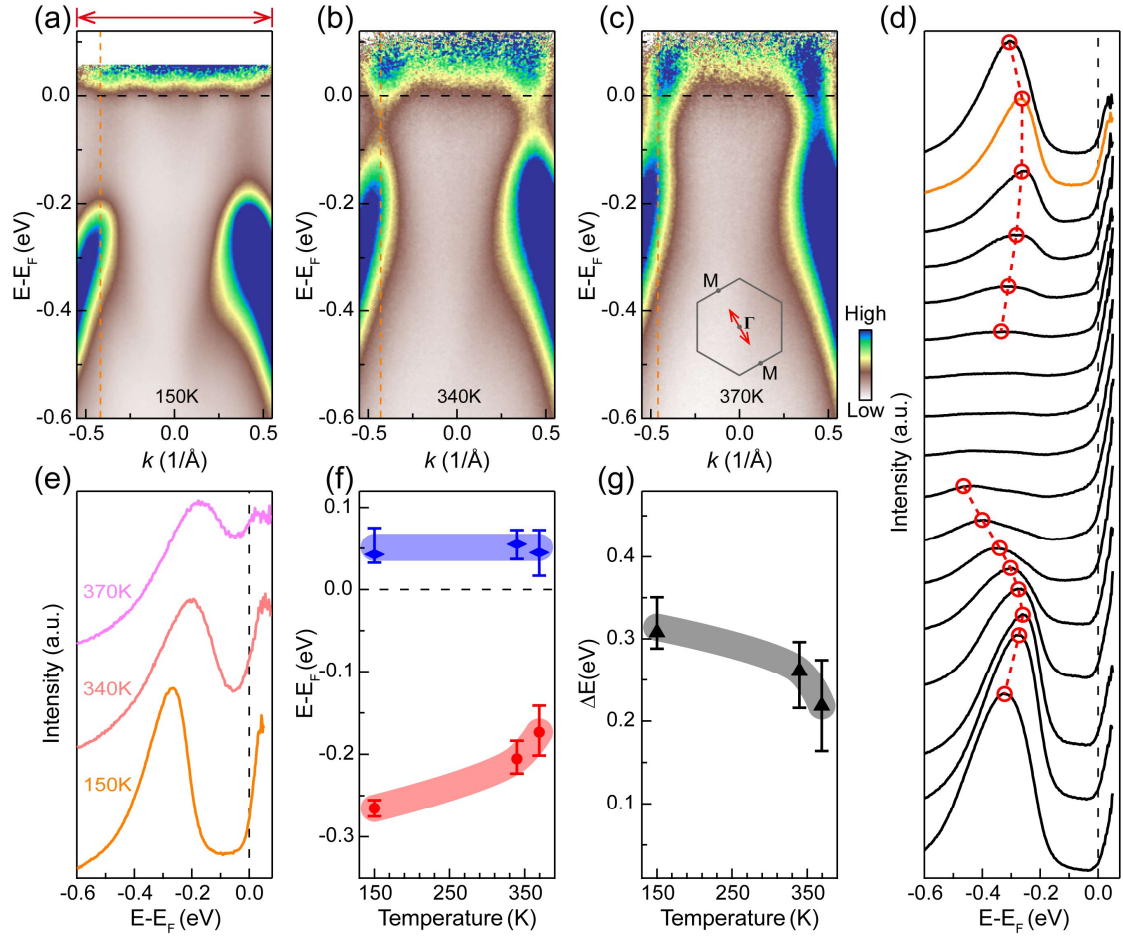

**Fig. 3 Temperature evolution of the electronic structure on the 1T termination. a-c,** Fermi-Dirac divided photoemission intensity plot of the band structure along M-Γ-M direction, measured at 150K, 340K, and 370K, respectively. The momentum location of the cut is shown in the inset of **c**. The yellow dashed line marks the momentum of the valence band top. **d,** Raw EDCs from **a**. The EDC across the valence band top is shown in yellow. Red circles mark EDC peaks of the hole-like valence band. **e,** Temperature evolution of the EDC across the valence band top. **f,** Temperature evolution of the valence band top (red circles) and the flat band (blue diamonds), extracted from **e**. Error bars represent the uncertainties in the determinations of energy positions for the valence band top and the flat band. **g,** Temperature evolution of the energy difference between the flat band and the valence band top. Error bars represent the uncertainties in the determination of the energy difference.

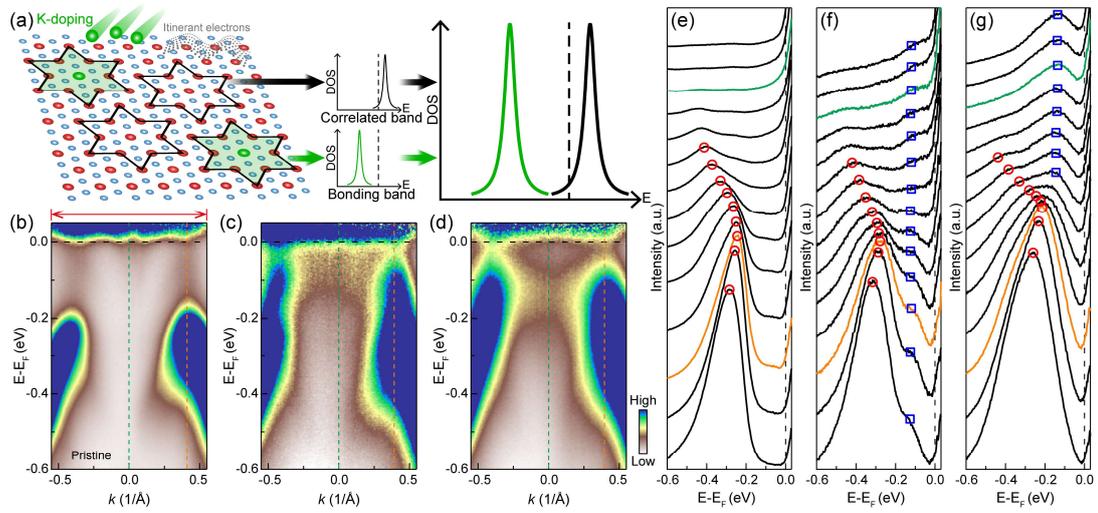

**Fig. 4 Evolution of the electronic structure with K surface doping on the 1T termination at 150K. a,** Schematic of K surface doping and the two-component local picture. **b,** Fermi-Dirac divided photoemission intensity plot of the band structure along M-Γ-M direction. The momentum location of the cut is shown in the inset of Fig. 3c. **c,d,** Same as **b**, but with continuous K surface doping. The green and yellow dashed lines indicate the Γ point and the momentum position of the valence band top. **e-g,** Raw EDCs from **b-d**, respectively. The EDC at Γ is shown in green, and the EDC across the valence band top is shown in yellow. Red circles mark the EDC peaks of the hole-like valence band and blue squares indicate the nondispersive new band.